\documentclass[prl,twocolumn,showpacs,nobibnotes,floatfix,superscriptaddress]{revtex4}

\usepackage{graphicx,eucal,amsfonts}

\begin{document}

\title{The Effects of Symmetries on Quantum Fidelity Decay}
\author{Yaakov S. Weinstein}
\thanks{To whom correspondence should be addressed}
\email{weinstei@dave.nrl.navy.mil}
\author{C. Stephen Hellberg}
\email{hellberg@dave.nrl.navy.mil}
\affiliation{Center for Computational Materials Science, Naval Research Laboratory, Washington, DC 20375 \bigskip}

\begin{abstract}
We explore the effect of a system's symmetries on fidelity decay behavior. 
Chaos-like exponential fidelity decay behavior occurs in non-chaotic systems 
when the system possesses symmetries and the applied perturbation is not tied 
to a classical parameter. Similar systems without symmetries exhibit 
faster-than-exponential decay under the same type of perturbation. This 
counter-intuitive result, that extra symmetries cause the system to behave in a
chaotic fashion, may have important ramifications for quantum error correction.
\end{abstract}

   
\maketitle
Fidelity decay as a possible signature of quantum chaos was introduced by 
Peres \cite{Peres1,Peres2} while exploring irreversibility in quantum 
mechanical systems. Though the overlap between two initial states undergoing 
equivalent evolution remains constant with time, it decreases with time if 
the Hamiltonian effecting one of the systems is sightly perturbed. The behavior
of the decrease in overlap, or fidelity, depends on whether the evolution is 
the quantum analog of a chaotic or non-chaotic classical system.

Further explorations have distinguished realms of fidelity decay behavior 
based on the chaoticity of the corresponding classical system, perturbation 
strength and type, and the initial system state. Weak perturbations of the 
Hamiltonian, such that perturbation theory is valid, exhibit a Gaussian 
fidelity decay \cite{Peres1,J1}. Stronger perturbations, in the Fermi golden
rule (FGR) regime, exhibit exponential fidelity decay with a rate determined 
by the strength and type of perturbation \cite{J1,P1,J2,Jo}. The rate of this 
exponential generally increases as the square of the perturbation strength 
\cite{J1} and may saturate at the underlying classical systems' Lyapunov 
exponent \cite{Jala,C1} or the bandwidth of the Hamiltonian \cite{J1}. 

The fidelity decay of classical-like coherent states for quantum analogs of 
non-chaotic systems \cite{P1,P3,L1} can be Gaussian \cite{P1,P2}, faster than 
the chaotic system exponential, or power-law \cite{J3} depending on the 
initial state \cite{P3} and the effect of the perturbation on the classical 
orbits on which the state is centered \cite{Ben2,YSW4}. The fidelity decay of 
systems \cite{Wang1}, states \cite{YSW2}, and perturbation strengths 
\cite{Cerr1} at regime edges have also been explored.

The quantum fidelity decay of an initial state $|\psi_i\rangle$ is
\begin{equation}
F(t) = |\langle\psi_i|U^{-t}U_p^t|\psi_i\rangle|^2
\end{equation}
where $U$ is the unperturbed evolution, $U_p = Ue^{-i\epsilon V}$ is the 
perturbed evolution, $\epsilon$ is the perturbation strength, and $V$ is the 
perturbation Hamiltonian. For chaotic systems, random matrix theory (RMT) 
gives the FGR exponential fidelity decay rate $\Gamma$ as a function of 
$\epsilon$ and the perturbation Hamiltonian eigenvalues $\lambda_i$ \cite{Jo}: 
\begin{equation}
\label{RMT}
\Gamma = \epsilon^2\overline{\lambda^2}, 
\end{equation}
where $\overline{\lambda^2} = N^{-1}\sum^N_i\lambda_i^2$ and $N$ is the Hilbert
space dimension.

Fidelity decay is related by a Fourier transform to the local density of 
states (LDOS) \cite{J1,W3} $\eta(\Delta\phi) = |\langle v_m|v'_n\rangle|^2$,
where $\Delta\phi = \phi_m-\phi'_n$, is the difference between unperturbed and 
perturbed eigenangles given by the eigenvalue equations: 
$U|v_m\rangle = \exp(-i\phi_m)|v_m\rangle$ and 
$U_p|v'_n\rangle = \exp(-i\phi'_n)|v'_n\rangle$. The LDOS provides a measure 
of how local is the perturbation. For complex systems the LDOS is typically 
Lorentzian \cite{LLDOS}, the perturbation transfers probability to 
far reaches of the system basis. The Lorentzian width $\Gamma$ gives the 
exponential chaotic fidelity decay rate. 

Recently Rossini, Benenti, and Casati (RBC) \cite{RBC} reported numerical 
evidence of exponential fidelity decay in non-chaotic systems due to `quantum' 
perturbations (though the rate may not be that expected for chaotic systems). 
Quantum perturbations are not tied to a classical system parameter and are 
applied multiple times during each map iteration, i.e. after each basic gate 
in the simulation of the dynamics on a quantum computer. The cause of this 
behavior, RBC explain, is the non-locality of such errors which allow for 
direct transfer of probability over a large distance in phase space. Meaning, 
the LDOS has significant amplitude even for large $\Delta\phi$. This result is 
of particular importance for suggested exploitations of the slower exponential 
decay to stabilize quantum computation \cite{P4,Shep1}, and for fidelity decay 
studies on a quantum computer \cite{Jo,RBC,Poulin,baker}. 

In this paper we numerically demonstrate that perturbations not tied to a 
classical parameter of non-chaotic systems can exhibit exponential decay 
even when applied after every map iteration, as generally done in fidelity 
decay studies. We choose this perturbation application scheme because 1) there 
is no unique gate sequence for implementing an operator 2) some of the 
operators we discuss cannot be implemented efficiently on a quantum computer
3) perhaps most importantly, when the perturbation is applied at numerous 
points during the map iteration the composite effect on the map is likely 
random. Thus, the exponential decay is due to the basis of the perturbation 
as found in \cite{Jo}. By applying the perturbation every map iteration we 
show that the decay behavior is due to a different phenomenon: symmetries in 
the system Hamiltonian. When no symmetries are present the perturbation 
is local, the LDOS is Gaussian, and the fidelity decay is faster than 
exponential even when the perturbation is not tied to a classical system 
parameter. We emphasize the counter-intuitive nature of this claim, namely, 
by adding \emph{symmetry}, the system's fidelity behavior behaves in a more 
\emph{chaotic} fashion. 

Our numerical study centers around the quantum kicked top (QKT), a system used 
in many studies of quantum chaos in general \cite{H1} and fidelity decay in 
particular \cite{Peres2,J1,P1,Jo}. The QKT \cite{H2} Floquet operator is
\begin{equation}
U_{QKT} = e^{-i\pi J_y/2}e^{-ik J_z^2/2J},
\end{equation}
where $J$ is the angular momentum of the top and ${\bf \vec{J}}$ are
irreducible angular momentum operators. The Hilbert space dimension of
the top is $N = 2J+1$ and the representation is such that $J_z$ is diagonal. 
The chaoticity of the QKT depends on the kick strength, $k$. The QKT is 
non-chaotic for $k \alt 2.7$, has chaotic and non-chaotic regions for 
$2.7 \alt k \alt 4.2$, and is fully chaotic for $k \agt 4.2$ \cite{J1}.  
The QKT has different symmetry sectors based on its angular momentum. For 
all $J$ the QKT has conserved parity with respect to $180^{\circ}$ 
rotations about $y$. For even $J$ the subspace even with respect to $y$ has 
conserved parity with respect to $180^{\circ}$ rotations about $x$.

In his original fidelity decay studies, Peres was careful to account for the
symmetries of the QKT \cite{Peres2} by block diagonalizing and performing
simulations using only one block. While some later authors have done the same 
\cite{P1}, others \cite{J1,YSW4} have used the complete QKT without apparent 
discrepancies. Here we show that the presence of symmetries in a quantum 
system can dramatically change the fidelity decay behavior.

We choose a `quantum' perturbation relevant to quantum control studies
\begin{equation}
\label{pert}
U_p = \Pi_{j=1}^{n_q} \exp(- i \epsilon \sigma_z^j / 2 ),
\end{equation}
where $n_q$ is the number of qubits and $\sigma_z$ is the Pauli spin matrix. 
This corresponds to a collective qubit rotation about the $z$-axis by
angle $\epsilon$ and is a model of coherent far-field errors \cite{Fortunato}. 
We apply the perturbation only after a complete map iteration.

\begin{figure}
\includegraphics[height=5.8cm, width=8cm]{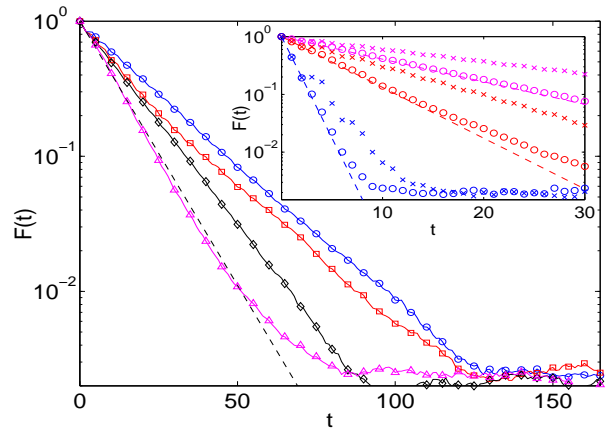}
\caption{\label{QKT}
(Color online) Fidelity decay for QKT with $J = 255.5$, and $k = 1$ ($\circ$), 
3 ($\Box$), 5 ($\diamond$), and 7 ($\triangle$) for the collective bit $z$ 
rotation with $\epsilon = .2$. Even the non-chaotic QKT decays are exponential
though they do not follow the random matrix theory rate given in Eq. 
\protect\ref{RMT} (dashed line). The inset shows the fidelity decay for 
$k = 1$ ($\times$) and 10 ($\circ$) for perturbation strengths 
$\epsilon = .2$, .3, and .6 (top to bottom). As the perturbation strengthens, 
the invariant subspaces are overwhelmed and the system behaves as if there were
no symmetries. Hence, the non-chaotic QKT reverts to non-exponential 
fidelity decay. RMT predictions are also shown (dashed lines). All plots are 
averaged over 100 random initial states.
}
\end{figure}

Figure \ref{QKT} shows the fidelity decay of the QKT for different values
of $k$ with the above perturbation, $\epsilon = .2$, and random initial states.
Values of $k$ are shown for regular, mixed, and chaotic QKT but the fidelity 
decay is always exponential. This tells us that the fidelity decay 
due to a quantum error is exponential independent of the chaoticity of the 
system \emph{even} when applied after every map iteration. Here there is no 
concern that the system has random eigenvectors in the basis of the 
perturbation since the perturbation is diagonal (and in this basis the regular 
QKT eigenvectors are not random). Initial angular momentum coherent states for 
the QKT and random states for the quantum Harper's map give similar results. 

\begin{figure}
\includegraphics[height=5.8cm, width=8cm]{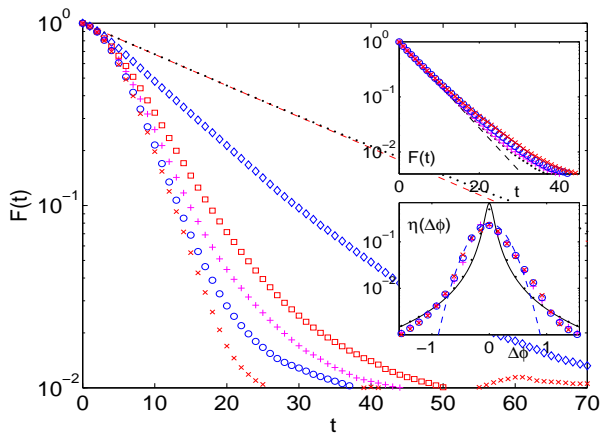}
\caption{\label{CUEd}
(Color online) Fidelity decay for matrices of the interpolating ensemble 
$\delta = 1$ ($\cdot$), .98 ($\diamond$), .94 ($\Box$), .9 (+), .8 ($\circ$), 
.7 ($\times$) for the collective qubit $z$ rotation with $N = 128$ and 
$\epsilon = .3$. The fidelity for $\delta < .7$ matrices has the same behavior 
as those with $\delta = .7$. For $\delta \neq 1$ the fidelity decay is faster 
than exponential even though the perturbation is not tied to a classical 
parameter (the matrices have no classical analogs). As $\delta \rightarrow 1$ 
the matrices become random and the decay rate approaches the RMT rate (dashed 
line). All plots average over 10 operators with 100 random initial states per 
operator. The lower inset shows the local density of states for an average of 
50 $\delta = 1$, .9, .8, .7 operators, compared to the Lorentzian expected for 
chaotic systems (solid line), and a Gaussian (dashed line). The Gaussian like 
LDOS for non-chaotic systems shows that the perturbation is localized. 
The upper inset shows the fidelity decay for $\epsilon = .3$, $N = 256$ block 
diagonal operators in which the two $N/2 = 128$ blocks are different 
interpolating ensemble matrices of the same $\delta$. Changing bases and 
having the perturbation break the symmetry of the operator, brings the 
fidelity decay close to the RMT prediction (dashed line). As $\delta$ is 
decreased, the fidelity decay becomes slightly slower than exponential.
}
\end{figure}

Exponential fidelity decay due to quantum perturbations is not universal. 
We demonstrate this via unitary matrices of the interpolating 
ensembles \cite{Zyc1}, which are intermediate between random, $\delta = 1$, 
and diagonal with Poissonian distributed eigenangles $\delta = 0$. Matrix 
statistics for these ensembles transition smoothly between the two limits 
\cite{Zyc1,YSW5}. There is no known way of efficiently implementing 
such matrices on a quantum computer. For our purposes matrices with 
$\delta < 1$ provide models of non-chaotic operators that have no classical 
analog. Any perturbation of these operators must be quantum as, a priori, 
the operators have no classical parameters. Fig. \ref{CUEd} demonstrates
that the $\delta \neq 1$ fidelity decay for the perturbation of Eq. 
\ref{pert}, $\epsilon = .3$, is \emph{not} exponential. 

An explanation for the above discrepancies is the presence of symmetries. 
Non-chaotic systems containing symmetries, and thus invariant subspaces, 
exhibit chaos-like exponential fidelity decay, while non-chaotic systems 
without symmetries exhibit faster-than-exponential decay. This result is 
surprising. A priori we would expect systems with less symmetries to be more 
chaos-like. Below we provide further numerical evidence demonstrating this 
behavior: one QKT symmetry sector and interpolating ensemble matrices with 
added symmetries.

\begin{figure}
\includegraphics[height=5.8cm, width=8cm]{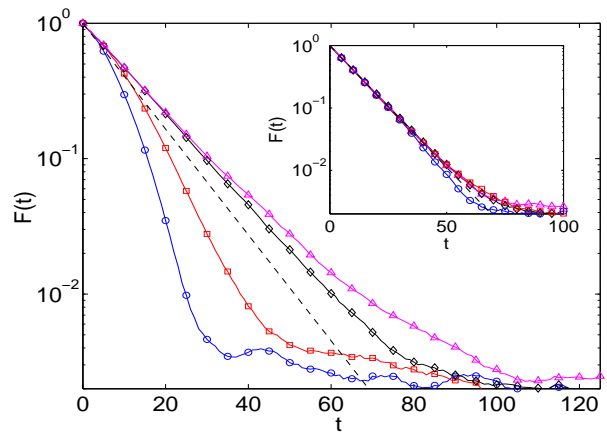}
\caption{\label{Sub}
(Color online) Fidelity decay for the QKT subspace even with respect to 
$180^{\circ}$ rotations about $y$ and odd with respect to $180^{\circ}$
rotations about $x$, $J = 1024$, $N = 512$, $k = 1$ ($\circ$), 3 ($\Box$), 
5 ($\diamond$), and 7 ($\triangle$) for the collective bit $z$ rotation with 
$\epsilon = .2$. Though the perturbation is not attached to a classical 
parameter the fidelity decay for non-chaotic QKTs are non-exponential. The 
chaotic QKT exponential decay does not follow the RMT rate, $\Gamma = .45$ 
(dashed line, actual rate is $\Gamma \simeq .405$). The inset shows the 
fidelity decay for the same parameters with the perturbation transformed into 
a random basis. The decay is exponential for all values of $k$ at the RMT rate.
}
\end{figure}

Following Peres \cite{Peres2} we block diagonalize an even $J$ QKT and keep 
one block as the system. Using the perturbation of Eq. \ref{pert}, 
the fidelity decay for the block is faster-than-exponential for 
non-chaotic values of $k$ and exponential for chaotic values of $k$, as shown 
in Fig. \ref{Sub}. This is in contrast to the exponential decay seen for all
$k$ values when using the complete QKT. Thus, by \emph{removing} the symmetry, 
the system deviates from the expected RMT behavior. In addition, for chaotic 
values of $k$, the exponential decay rate is not the one given by Eq. 
\ref{RMT}, again in contrast to the full QKT where the decay rate was exact.

We expect the fidelity decay of systems with symmetries and extremely strong 
perturbations, such that the symmetries are overwhelmed, to revert to 
non-exponential. This occurs in the full QKT with $\epsilon = .6$, Fig. 
\ref{QKT} inset.

A symmetry can be added to interpolating ensemble matrices by using two 
matrices of equal $\delta$ and Hilbert dimension $N/2$ as diagonal blocks 
of a $N\times N$ operator. The perturbation of Eq. \ref{pert} does not cause 
mixing between the blocks so we transform into a new basis. The first 
transformation we apply takes the perturbation to a collective qubit $x$ 
rotation, $\sigma_x$ replaces $\sigma_z$ in Eq. \ref{pert}. With this the 
fidelity decay is exponential at the RMT rate even for low $\delta$ 
blocks. However, this exponential decay is not due to the broken symmetry. 
Rather, the matrices of the interpolating ensembles, though not random with 
respect to the $\sigma_z$ collective qubit perturbation, are random in the 
eigenbasis of the $\sigma_x$ collective qubit perturbation. As explained in 
\cite{Jo} (and shown in Fig. \ref{Sub} for the QKT without symmetry), when the 
system is random in the perturbation basis, the fidelity decay is exponential. 

A transformation which does not induce randomness in a block diagonal operator
with interpolating ensemble blocks is a modified version of the 
transformation matrix to block diagonalize the even $J$ QKT \cite{Peres2,BD}. 
With this transformation the fidelity decay is exponential for $\delta \neq 1$ 
blocks and becomes slightly slower than exponential for lower values of 
$\delta$, Fig. \ref{CUEd}. Thus, by \emph{adding} a symmetry to the system 
we have slowed the fidelity decay and made its behavior more chaos-like.

The above simulations demonstrate that perturbations of non-chaotic systems, 
even ones not attached to a classical system parameter, tend to be local. 
However, when the perturbation breaks a symmetry there are long range effects 
similar to those of chaotic systems. This process is different than that 
described in Ref. \cite{Jo} where the exponential decay is linked to the RMT 
statistics of the system eigenvectors in the perturbation basis. 

Coherent far-field errors, such as Eq. \ref{pert}, are the subject of many 
theoretical and experimental quantum error-correction codes and encodings 
\cite{Fortunato,FFN}. Our results suggest that this error may cause only 
exponential fidelity decay even when a quantum computer is simulating 
non-chaotic evolution. This provides a novel way of protecting against these 
noise operators, \emph{add a symmetry to the system}. Such symmetries already 
exist in encoded qubits, where specific states of a multi-qubit system are 
used as a logical qubit. For example, logical qubits in quantum dots that are 
encoded such that the exchange coupling is universal \cite{Div,YSW6} are 
symmetric with respect to the $S_z$ angular momentum operator. In general, by 
enforcing all reasonable decoherence mechanisms to break a symmetry, 
decoherence may cause an exponential, rather than Gaussian decay. 

Exponential decay occurs in many physical systems not generally regarded as 
chaotic, including relaxation phenomena and Fermi Golden Rule calculations. 
We suggest, without proof, that symmetry may add insight into the preponderance
of exponential decay laws in nature. Lack of symmetry may explain why some 
systems, such as magnetic relaxation in single molecular magnets \cite{SMM}, 
decay non-exponentially. 

In conclusion, we have studied the effect of symmetries on fidelity decay
behavior. When the perturbation is not tied to a classical parameter of the 
system, as would likely arise in quantum computers, the presence or lack 
of symmetries strongly affects the fidelity decay behavior. Surprisingly, 
the presence of symmetries forces the system into a more chaos-like behavior, 
exponential decay,  while lack of symmetries causes deviations from RMT 
predictions and a faster-than-exponential decay. Building symmetries 
into quantum computers, as done when encoded qubits, can cause decoherent 
processes to affect the system with the less-damaging exponential decay. 

The authors acknowledge support from the DARPA QuIST (MIPR 02 N699-00) 
program. Y.S.W. acknowledges support of the National Research Council 
through the Naval Research Laboratory. Computations were performed at the 
ASC DoD Major Shared Resource Center.


\begin{thebibliography}{99}
\bibitem{BD}
The even $J$ QKT transformation matrix has odd Hilbert space dimension $2J+1$. 
To make $N$ even we remove the eigenvector that is all zeros except a one at 
the $N/2+1$ element. The $N/2+1$ element in all other eigenvectors is zero 
and is removed to give them a length of $N$.
\end{thebibliography}
\end{document}